\documentclass[amstex,overcite,amsfonts,aps,prb,twocolumn,amsmath]{revtex4}
\usepackage{graphicx}
\begin{document}
\title{\textbf{Polarization and Experimental Configuration Analysis of Sum Frequency Generation
Vibrational Spectra of Air/Water Interface}}
\author{\textsf{Wei Gan}\footnote[2]{Also Graduate School of the Chinese
Academy of Sciences}}
\author{\textsf{Dan Wu\footnotemark[2]}}
\author{\textsf{Zhen Zhang\footnotemark[2]}}
\author{\textsf{Ran-ran Feng\footnotemark[2]}}
\author{\textsf{Hong-fei Wang}\footnote[1]{Author to whom correspondence should be
addressed. E-mail: hongfei@mrdlab.icas.ac.cn. Tel. 86-10-62555347,
Fax 86-10-62563167.}}\affiliation{State Key Laboratory of
Molecular Reaction Dynamics,
\\Institute of Chemistry, the Chinese Academy of Sciences,
Beijing, China, 100080}

\date{\today}

\begin{abstract}
Here we report a detailed study on spectroscopy, structure and
dynamics of water molecules at air/water interface, investigated
with Sum Frequency Generation Vibrational Spectroscopy (SFG-VS).
Quantitative polarization and experimental configuration analysis
of the SFG data in different polarizations with four sets of
experimental configurations can shed new lights on our present
understanding of the air/water interface. Firstly, we concluded
that the motion of the interfacial water molecules can only be in
a limited angular range, instead rapidly varying over a broad
angular range in the vibrational relaxation time suggested
previously. Secondly, because different vibrational modes of
different molecular species at the interface has different
symmetry properties, polarization and symmetry analysis of the
SFG-VS spectral features can help assignment of the SFG-VS spectra
peaks to different interfacial species. These analysis concluded
that the narrow $3693cm^{-1}$ and broad $3550cm^{-1}$ peaks belong
to $C_{\infty v}$ symmetry, while the broad $3250cm^{-1}$ and
$3450cm^{-1}$ peaks belong to the symmetric stretching modes with
$C_{2v}$ symmetry. Thus, the $3693cm^{-1}$ peak is assigned to the
free OH, the $3550cm^{-1}$ peak is assigned to the single hydrogen
bonded OH stretching mode, and the $3250cm^{-1}$ and $3450cm^{-1}$
peaks are assigned to interfacial water molecules as two hydrogen
donors for hydrogen bonding (with $C_{2v}$ symmetry),
respectively. Thirdly, analysis of the SFG-VS spectra concluded
that the singly hydrogen bonded water molecules at the air/water
interface have their dipole vector direct almost parallel to the
interface, and is with a very narrow orientational distribution.
The doubly hydrogen bond donor water molecules have their dipole
vector point away from the liquid phase.
\end{abstract}

\maketitle

\section{Introduction}

Interfaces of water are the most important subjects not only
because water is widely involved in physical, chemical,
environmental as well as biological processes, but also because
water is so far the most mysteries molecule in the
universe.\cite{FranksBook,science-Luecke,TobiasPRL2002,RicePNAS1999}
Among them, air/water interface has been intensively investigated
theoretically or experimentally over the last decades.
Spectroscopy, molecular structure and dynamics at air/water
interface is studied with theoretical analysis such as \textit{ab
initio} calculation or molecular dynamics
simulation,\cite{Mundy-science,
BenjaminPRL1994,HynesCP2000,HynesJPCB2002,MooreJCP2003,ChandraCPL2003,
ChandraCPL2004,RiceJCP1991,BenjaminCR1996} or experimental
techniques such as X-ray
reflection,\cite{PershanPRL1985,PershanPRA1988} Stimulated Raman
Scattering (SRS),\cite{SawadaCPL2002} Near-edge X-Ray Adsorption
Fine Structure (NEXAFS),\cite{SaykallyJPCM2002} Second Harmonic
Generation (SHG),\cite{EisenthalJPC1988,FreyMP2001} as well as Sum
Frequency Generation, etc.\cite{ShultzIRPC2000,
RichmondARPC2001,Richmond:cr102:2693,Shen-science,richmond:science,Shen-prl1994,
DuQuanPRL1993,Richmond-jpca2000,WeiXingPRL2001} Among these
experimental techniques, Second Harmonic Generation and Sum
Frequency Generation are the most important methods for molecular
interface studies because of their surface sensitivity and
specificity.\cite{RichmondSHGReview,ShenANRP1989,CornHigginsReview1994,
ShenMirandaJPCBReview,shen:nature:review,eisenthal:review,Shen-5CT,somorjai:ap:review}
With these investigations, the properties of the water molecules
at the interface, such as the surface density, surface structure,
surface potential as well as surface dynamics, have been
intensively discussed.

However, conclusions on the surface molecule species at air/water
interface are still under discussion.\cite{DuQuanPRL1993,
Richmond-jpca2000,WeiXingPRL2001,SaykallyJPCM2002} With SFG-VS
experimental studies, the following interfacial water species have
been reported in literatures, namely, water molecules straddle at
the interface with one OH bond hydrogen bonded to neighboring
molecules in liquid phase (singly bonded OH) and another OH bond
free from hydrogen bonding (free OH) in gas
phase;\cite{Shen-prl1994,ShultzIRPC2000,Richmond-jpca2000,richmond:jpcb1998}
water molecules with both OH bonds symmetrically hydrogen bonded
in a tetrahedral network (ice-like and liquid-like
structures);\cite{Shen-prl1994,
ShultzIRPC2000,Richmond-jpca2000,richmond:jpcb1998} and water
molecules in gas phase with both OH bonds not hydrogen bonded
pointing into the liquid
phase.\cite{Richmond-jpca2000,richmond:science} With NEXAFS
measurement and \textit{ab initio} molecular dynamics simulation,
water molecules with both OH bonds not hydrogen bonded pointing
out of the interface was also
proposed.\cite{SaykallyJPCM2002,Mundy-science} The latter case is
particularly controversial because NEXAFS is not strictly a
surface specific technique.\cite{SaykallyJPCM2002} With
polarization SFG-VS measurement, Wei \textit{et al.} discussed the
absence of SFG spectra in some polarization combinations and
proposed an explanation through fast orientational motion in a
broad range of about $102^{\circ}$ in a time scale comparable or
less than 0.5 \textit{ps}.\cite{WeiXingPRL2001} However, puzzle
still remains because some of the experimental studies suggests
ordered and slow dynamics for interfaces of hydrogen bonding
liquids, while some experimental investigations suggested a more
dynamic and less ordered picture for the liquid interfaces,
air/water interface included.\cite{ShenMirandaJPCBReview} In
addition, whether the surface orientation relaxation is fast or
slow than the bulk water molecules is also an issue under
discussion in the recent
literatures.\cite{BakkerScience,ChandraCPL2004,BenjaminJPCBASAP}
Besides SHG and SFG-VS experimental
studies,\cite{Eisenthal1992ACR,ShenMirandaJPCBReview} Structure
and dynamics of water molecules at the air/water interface have
also been intensively discussed with theoretical
simulations.\cite{Mundy-science,BenjaminPRL1994,HynesCP2000,HynesJPCB2002,
ChandraCPL2003,ChandraCPL2004,MooreJCP2003,RiceJCP1991,BenjaminCR1996}
Even though with so much efforts and progresses both by
experimentalists and theoreticians, our detailed understanding of
air/water interface is still limited. Just as indicated by B. C.
Garrett recently,\cite{GarrettScience} `...(direct) experiments
are difficult to perform because the liquid interface is
disordered, dynamic, and small (typically only a few molecules
wide) relative to the bulk'.

Actually, direct measurement of the liquid interface is not as
difficult as suggested as above. It has been known that along with
SHG, SFG-VS can provide direct measurement on liquid interface no
other technique can
match.\cite{Eisenthal1992ACR,ShenMirandaJPCBReview} As pointed out
by Miranda and Shen, `SFG is currently the only technique that can
yield a vibrational spectrum for a neat liquid
interface'.\cite{ShenMirandaJPCBReview} In fact, with the advances
of ultrafast laser and detection technology in the past decade and
especially recent few
years,\cite{RichterOL1998,AllenAnaSci2001BroadBandSFG,RichmondApplySpec2004,
JohnsonPCCP2005} particularly with commercial systems designed for
SFG-VS measurement,\cite{EKSPLA&EROSACAN} SFG-VS, as well as SHG,
experiments have come from easier to routine.\cite{ShenApplyPhys}
The real difficulty lies on the fact that quantitative analysis
and interpretation of the SFG-VS, as well as SHG, data had been
not as well developed and widely performed until
recently.\cite{Shen-5CT,WeiXingPRL2001,weixing:pre2000,WHFRaoJCP2003,Lurong1,Lurong2,
HongfeiCJCPPaper,Lurong3,ChenhuaJPCBacetone,
ChenhuaJPCBmethanol,ChenhuaCPLacetone,GanweiCPLNull,HongfeiIRPCreview}
Therefore, conclusions in many previous reports on the
investigations of air/water interface, as well as other liquid
interfaces, with SFG-VS are subjected to different
interpretations.

As we have demonstrated in a series of recent publications,
systematically quantitative treatment to SFG-VS data is not only
possible, but also very effective for obtaining detailed
spectroscopic, structural and thermodynamic properties of liquid
interfaces.\cite{WHFRaoJCP2003,Lurong1,Lurong2,
HongfeiCJCPPaper,Lurong3,ChenhuaJPCBacetone,ChenhuaJPCBmethanol,
ChenhuaCPLacetone,GanweiCPLNull,HongfeiIRPCreview} In these works,
we not only developed methodology for quantitative polarization
and experimental configuration analysis in SFG-VS and SHG, we also
tested accuracy and sensitivity of some of the methodology. We
have applied them to elucidated the anti-parallel double layered
structure and thermodynamics of some organic liquid aqueous
solution interfaces. In addition, we also demonstrated that a set
of polarization selection rules (or guidelines) in SFG-VS can be
developed for vibrational spectrum assignment through symmetry
analysis of the SFG-VS spectral features.\cite{Lurong2,Lurong3}
This latter approach is extremely useful for discerning complex
SFG-VS spectrum with unidentified or controversial assignments.
Recently, based on polarization analysis, Ostroverkhov \textit{et
al.} demonstrated a phase-sensitive interference analysis of SFG
polarization spectra of water/quartz
interface.\cite{Shen2005PRLWaterQuartz} With these development, in
this report we intend to apply these analysis methodologies to the
study of air/water interface.

In this work, we examined SFG-VS spectra at air/water interface
measured in different polarizations under four experimental
configurations with polarization analysis method and experimental
configuration analysis. With these analysis, detailed new
information are obtained for understanding of the spectroscopy,
structure and dynamics of the air/water interface. In the
following sections, after a brief introduction of the theoretical
background and experimental conditions, we first discuss the
motion of the interfacial water molecules at the air/water
interface, which was previously suggested experiencing rapidly
motion over a broad angular range in the vibrational relaxation
time; then we use polarization and symmetry analysis of the SFG-VS
spectral features for assignment of the SFG-VS spectra peaks; in
the end, we shall discuss the structure and orientation of the
water molecules at the air/water interface.

\section{Polarization and Experimental Configuration Analysis in SFG-VS}

Quantitative polarization analysis and experimental configuration
analysis can provide rich and detailed information of
spectroscopy, structure and dynamics of molecular
interfaces.\cite{Shen-5CT,WeiXingPRL2001,WHFRaoJCP2003,Lurong2,Lurong3,GanweiCPLNull}
Generally, the SFG intensity in the reflective direction
is,\cite{Lurong2,Shen-5CT}

\begin{eqnarray}
I(\omega)&=&\frac{{8\pi ^3 \omega ^2sec^2\beta
}}{{c^{3}n_{1}(\omega)n_{1}(\omega_{1}})n_{1}(\omega_{2})}\left|\chi
_{eff}^{(2)}\right|^2 I(\omega_{1})I(\omega_{2})\label{all}
\end{eqnarray}
\noindent

\noindent in which $\omega$, $\omega_{1}$ and $\omega_{2}$ are the
frequencies of the SFG signal, visible and IR laser beam,
respectively. $n_{i}(\omega_{i})$ is the refractive index of bulk
medium $i$ at frequency $\omega_{i}$, and $n'(\omega_{i})$ is the
effective refractive index of the interface layer at $\omega_{i}$.
$\beta_{i}$ is the incident or reflection angle from interface
normal of the $i$th light beams; $I(\omega_{i})$ is the intensity
of the SFG signal or the input laser beam. $\chi_{eff}^{(2)}$ is
the effective second order susceptibility for an interface. The
notations and the experiment geometry have been described in
detail previously.\cite{Lurong2,Shen-5CT}

$\chi_{eff}^{(2)}$ for the four generally used independent
polarization combinations can be deduced from the 7 nonzero
macroscopic susceptibility tensors for an achiral rotationally
isotropic interface ($C_{\infty v}$).\cite{Lurong2,Shen-5CT}

\begin{eqnarray}
\chi_{eff}^{(2),ssp}&=&
L_{yy}(\omega)L_{yy}(\omega_{1})L_{zz}(\omega_{2})sin\beta_{2}\chi_{yyz}\label{ssp}\nonumber
\\
\chi_{eff}^{(2),sps}&=&L_{yy}(\omega)L_{zz}(\omega_{1})L_{yy}(\omega_{2})sin\beta_{1}\chi_{yzy}\label{sps}\nonumber
\\
\chi_{eff}^{(2),pss}&=&L_{zz}(\omega)L_{yy}(\omega_{1})L_{yy}(\omega_{2})sin\beta\chi_{zyy}\label{pss}\nonumber
\\
\chi_{eff}^{(2),ppp}&=&
-L_{xx}(\omega)L_{xx}(\omega_{1})L_{zz}(\omega_{2})
cos\beta{cos\beta_{1}}sin\beta_{2}\chi_{xxz}\nonumber\\
&&-L_{xx}(\omega)L_{zz}(\omega_{1})L_{xx}(\omega_{2})cos\beta{sin\beta_{1}}cos\beta_{2}\chi_{xzx}\nonumber\\
&&+L_{zz}(\omega)L_{xx}(\omega_{1})L_{xx}(\omega_{2})sin\beta{cos\beta_{1}}cos\beta_{2}\chi_{zxx}\nonumber\\
&&+L_{zz}(\omega)L_{zz}(\omega_{1})L_{zz}(\omega_{2})sin\beta{sin\beta_{1}}sin\beta_{2}\chi_{zzz}\nonumber\\
\label{ppp}
\end{eqnarray}

\noindent It is so defined that the $xy$ plane in the laboratory
coordinates system $\lambda(x,y,z)$ is the plane of interface; all
the light beams propagate in the $xz$ plane; $\textit{p}$ denotes
the polarization of the optical field in the $xz$ plane, with $z$
as the surface normal, while $\textit{s}$ the polarization
perpendicular to the $xz$ plane. The consecutive superscript, such
as \textit{ssp}, represents the following polarization
combinations: SFG signal \textit{s} polarized, visible beam
\textit{s} polarized, IR beam \textit{p} polarized, and so forth.
$L_{ii}$ ($i=x,y,z$) is the Fresnel coefficient determined by the
refractive indexes of the two bulk phase and the interface layer,
and the incident and reflected angles.\cite{Lurong2,Shen-5CT}
$\chi_{ijk}^{(2)}$ tensors are related to the microscopic
hyperpolarizability tensor $\beta_{i'j'k'}^{(2)}$ of the molecules
in the molecular coordinates system $\lambda'(a,b,c)$ through the
ensemble average over all possible molecular orientations.
\cite{Lurong2,Shen-5CT}

\begin{eqnarray}
\chi^{(2)}_{ijk}&=&N_{s}\sum_{i'j'k'}\langle{R_{ii'}R_{jj'}R_{kk'}\rangle}\beta_{i'j'k'}^{(2)}
\label{hyper}
\end{eqnarray}

\noindent where $R_{\lambda\lambda'}(\theta,\phi,\psi)$ is the
matrix element of the Euler rotational transformation matrix from
the molecular coordination $\lambda'(a,b,c)$ to the laboratory
coordination $\lambda$(\textit{x,y,z}); $\beta_{i'j'k'}^{(2)}$ is
the microscopic (molecular) hyperpolarizability
tensor.\cite{GoldsteinBook,HongfeiCJCPPaper,HongfeiIRPCreview}
Here $N_{s}$ is the molecular number density at the interface.
$\langle A \rangle$ represents orientational average of property
$A(\theta,\phi,\psi)$ over the orientational distribution function
$f(\theta,\phi,\varphi)$.

\begin{eqnarray}
\langle A
\rangle=\frac{\int^{\pi}_{0}\int^{2\pi}_{0}\int^{2\pi}_{0}A(\theta,\phi,\psi)
f(\theta,\phi,\psi)\sin\theta d\theta\ d\phi d\psi}
{\int^{\pi}_{0}\int^{2\pi}_{0}\int^{2\pi}_{0}f(\theta,\phi,\psi)\sin\theta
d\theta d\phi d\psi}\label{OAverage}
\end{eqnarray}

For SFG-VS, $\beta^{(2)}$ is IR frequency ($\omega_{2}$)
dependent,

\begin{eqnarray}
\beta^{(2)}_{i'j'k'}&=&\beta_{NR,i'j'k'}^{(2)}+\sum_{q}\frac{\beta_{q,i'j'k'}}
{\omega_{2}-\omega_{q}+i\Gamma_{q}}\label{spectrum}
\end{eqnarray}

Thus, $\chi_{ijk}^{(2)}$ can be expressed into,

\begin{eqnarray}
\chi_{ijk}^{(2)}&=&\chi_{NR,ijk}^{(2)}+\sum_{q}\frac{\chi_{q,ijk}}
{\omega_{2}-\omega_{q}+i\Gamma_{q}}\label{spectrum1}
\end{eqnarray}

Therefore, SFG-VS measures the vibrational spectroscopy of
molecular interfaces. For dielectric interfaces, such as liquid
interfaces, the non-resonant term $\beta_{NR,i'j'k'}^{(2)}$ or
$\chi_{NR,ijk}^{(2)}$ is generally negligible compare with the
resonant terms.

Recently, we have found that the following formulation is very
effective in quantitative polarization and orientation analysis of
SFG and SHG data. It can be generally shown that in surface SFG
and SHG for an interface with orientational order, the effective
second order susceptibility $\chi_{eff}^{(2)}$ can be simplified
into the following form.\cite{WHFRaoJCP2003}

\begin{equation}
\chi_{eff}^{(2)} = N_{s}\ast\textit{d}\ast(\langle \cos \theta
\rangle - \textit{c}\ast \langle \cos ^3\theta \rangle
)=N_{s}\ast\textit{d}\ast \textit{r}(\theta) \label{chi}
\end{equation}

\noindent $r(\theta)$ is called the \textit{orientational field
functional}, which contains all molecular orientational
information at a given SFG experimental configuration; while the
dimensionless parameter \textit{c} is called the \textit{general
orientational parameter}, which determines the orientational
response $r(\theta)$ to the molecular orientation angle $\theta$;
and $\textit{d}$ is the susceptibility strength factor, which is a
constant in a certain experimental polarization configuration with
a given molecular system. The $d$ and $c$ values are both
functions of the related Fresnel coefficients including the
refractive index of the interface and the bulk phases, and the
experimental geometry.

The key for quantitative analysis is that both \textit{d} and
\textit{c} can be explicitly derived from the expressions of the
$\chi_{eff}^{(2)}$ in relationship to the macroscopic
susceptibility and microscopic (molecular) hyperpolarizability
tensors for a particular molecular vibrational
modes,\cite{Lurong2,HongfeiIRPCreview} as shown for the water
molecules with $C_{2v}$ symmetry in the appendix. With the
parameters $c$ and $d$, the polarization dependence and the
orientation dependence of the SFG/SHG signal for a certain
interface at certain experimental configuration can be analyzed
and calculated with clear physical picture on molecular
orientation and orientational distribution.\cite{WHFRaoJCP2003}
Reciprocally, information on the molecular symmetry, molecular
orientation and dynamics can be obtained from the analysis on the
SFG intensity relationships measured in different polarization
combinations and experimental
configurations.\cite{Lurong2,Lurong3,HongfeiIRPCreview,GanweiCPLNull}

The orientational average in Eq.\ref{hyper} is only the static
average on molecular orientations, without considering fast
molecular motion effects. Recently Wei \textit{et al.} discussed
the fast and slow limit of the time average over orientational
motion for $\chi_{eff}^{(2)}$, and they also applied this
treatment to analysis the polarization dependence of SFG
measurement of the OH stretching vibrational spectra for the
air/water interface.\cite{WeiXingPRL2001} In the fast motion
limit, the orientational motion is faster than the vibrational
relaxation time scale $1/\Gamma_{q}$ of the $q$th vibrational
mode; while in the slow motion limit, the orientational motion is
much slower than $1/\Gamma_{q}$.

According to Wei \textit{et al.},\cite{WeiXingPRL2001} the slow
motion limit gives,

\begin{eqnarray}
\chi^{(2)}_{ijk}&=&N_{s}\sum_{q}\sum_{i'j'k'}
\frac{\beta^{(2)}_{q,i'j'k'}}{\omega_{2}-\omega_{q}+i\Gamma_{q}}
\langle R_{ii'}R_{jj'}R_{kk'}\rangle \label{SlowAverage}
\end{eqnarray}

\noindent while the fast motion gives,

\begin{eqnarray}
\chi^{(2)}_{ijk}&=&N_{s}\sum_{q}\sum_{i'j'k'}
\frac{\beta^{(2)}_{q,i'j'k'}}{\omega_{2}-\omega_{q}+i\Gamma_{q}}
\langle R_{ii'}R_{jj'}\rangle \langle R_{kk'}
\rangle\label{FastAverage}
\end{eqnarray}

\noindent in which
$R_{\lambda\lambda'}(t)=\hat{\lambda}\cdot\hat{\lambda'}(t)$ is
the time-dependent direction Euler transformation matrix from
$\lambda'(a,b,c)$ to $\lambda(x,y,z)$ coordinates system. Because
of the molecular orientational motion, the molecular coordinates
$\lambda'(a,b,c)$ is time-dependent. Eq.\ref{SlowAverage} is
equivalent to Eq.\ref{spectrum1}, which is obtained by insertion
of Eq.\ref{spectrum} into Eq.\ref{hyper}.

\section{Experiment}

The details of the laser system has been described in our previous
reports.\cite{Lurong2,ChenhuaJPCBacetone,ChenhuaJPCBmethanol}
Briefly, the 10Hz and 23 picosecond SFG spectrometer laser system
(EKSPLA) is in a co-propagating configuration. The efficiency of
the detection system has been improved for the weak SFG signal of
air/water interface. A high-gain low-noise photomultiplier
(Hamamatsu, PMT-R585) and a two channel Boxcar average system
(Stanford Research Systems) are integrated into the EKSPLA system.
The voltage of R585 was 1300V in the measurement for air/water
interface, and 900V for the Z-cut quartz surface. The wavelength
of the visible is fixed at 532nm and the full range of the IR
tunability is $1000cm^{-1}$ to $4300cm^{-1}$. The specified
spectral resolution of this SFG spectrometer is $<6cm^{-1}$ in the
whole IR range, and about $2cm^{-1}$ around $3000cm^{-1}$. Each
scan was with a $5cm^{-1}$ increment and was averaged over 300
laser pulses per point. Each spectrum has been repeated for at
least several times. Moreover, for \textit{sps} polarization, each
spectrum has been repeated for more than a dozen times and
averaged. The energy of visible beam is typically less than
300$\mu J$ and that of IR beam less than 150$\mu J$ around
$3000cm^{-1}$ and $3700cm^{-1}$, and less than 100$\mu J$ in the
region in between. These are comparable to literature reported
values for measurement of air/water interface.\cite{DuQuanPRL1993}
All measurements were carried out at controlled room temperature
($22.0\pm0.5^{\circ}C$) and humidity (40$\%$) . The sample used
was ultrapure water from standard Millipore treatment (18.2
M$\Omega \cdot cm$). The whole experimental setup on the optical
table was covered in a plastic housing to reduce the air flow. No
detectable evaporation effect was observed for SFG spectrum during
each scan.

The normalization procedure of the SFG signal in different
experimental configurations need to be specifically discussed. The
detail of the normalization procedure for a single experimental
configuration was presented in Xing Wei's Ph.D.
dissertation.\cite{Wei:Thesis} However, the difference of coherent
length and Fresnel factors with different incident angles in the
quartz SFG signal measurements has to be corrected when comparing
SFG signal in different experimental configurations. Therefore,
the measured spectrum is firstly normalized with the energy of the
incident laser beams, and then normalized to the SFG signal of
Z-cut quartz (also normalized by the energy of the incident
lasers). Then it times with a converting factor between different
experimental configurations. This factor contains the influence of
the coherent length of Z-cut quartz,\cite{Wei:Thesis} the Fresnel
coefficients,\cite{Wei:Thesis} the $\chi_{ijk}$ value for Z-cut
quartz, and the factor $sec^{2}\beta$ for each experimental
configuration. Therefore, the end result is directly proportional
to the SFG intensity in Eq.\ref{all}. If the spectrum in Fig.
\ref{allSpectra} is divided by the factor $sec^{2}\beta$ and the
factor of the PMT efficiency between 1300V and 900V, which is
determined as 24.1 in our detection system, and then times the
unit factor $1\times 10^{-40}V^{4}m^{-2}$ which we left out for
simplicity of graph presentation, it will give the value for
$|\chi^{(2)}_{eff}|^{2}$. For example, the peak at about
$3700cm^{-1}$ in the \textit{ssp} spectra of Config.2 in
Fig.\ref{allSpectra} is about 0.23 unit. After above conversion it
gives $|\chi^{(2)}_{eff}|^{2}=4.7\times 10^{-40}V^{4}m^{-2}$,
matching satisfactorily with the reported value for less than
$10\%$ difference.\cite{WeiXingPRL2001}

Even though the normalized intensities are generally consistent
with each other, there can be possibly other sources of errors
when intensities in different experimental configurations need to
be compared. Because the visible and IR beams have different
coherent lengthes in the Z-cut crystal, and because these coherent
lengthes vary with different experimental incident angles, one of
the most likely error might come from the different focusing
parameters with different beam overlapping quality of the visible
and IR beams in the Z-cut quartz crystal with different
experimental configurations. Therefore, quantitative comparison of
the SFG spectral intensities in different polarizations with the
same experimental configuration can be more accurate than
comparison intensities between different experimental
configurations. Even though the latter is a good solution to
reduce such relative error associated with different experimental
configurations need to be developed.

\section{Results and Discussion}

\subsection{Polarization SFG Spectra of the air/water interface}

Firstly we would like to present the polarization SFG spectra of
the air/water interface measured in four different experimental
configurations.

\begin{figure*}[t]
\begin{center}
\includegraphics[height=15cm,width=15cm]{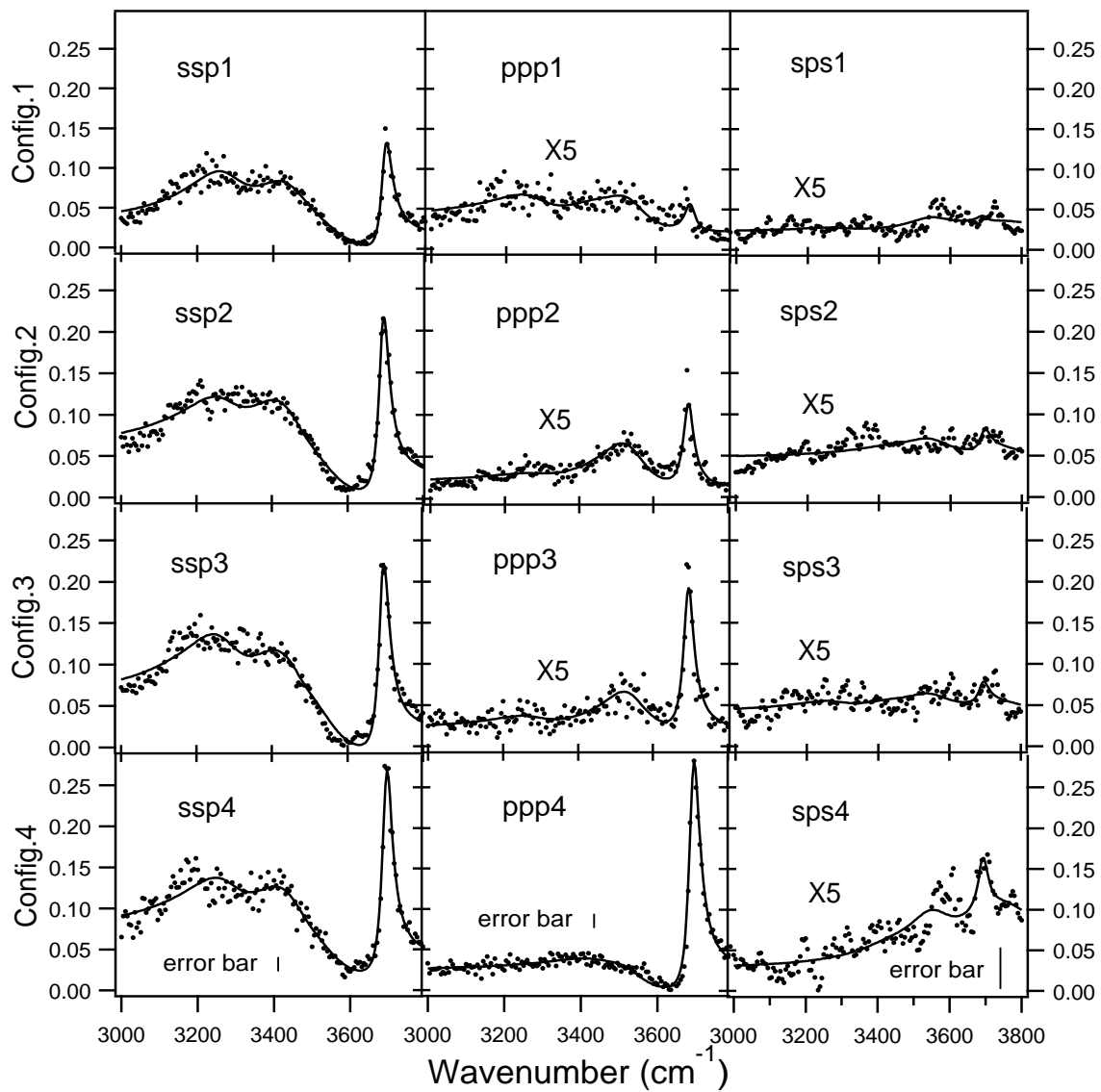}
\caption{SFG spectra of air/water interface in different
polarization combination and experimental configurations. All
spectra are normalized to the same scale. The solid lines are
globally fitted curves with Lorentzian line shape function in Eq.
\ref{spectrum1}. Note the different error bars for graphs in
different scales.}\label{allSpectra}
\end{center}
\end{figure*}

We have demonstrated recently that the change of the SFG spectra
in different polarizations by varying the experimental
configurations can be used for quantitative polarization analysis
and orientational analysis.\cite{HongfeiIRPCreview,GanweiCPLNull}
Here we present in Fig.\ref{allSpectra} the SFG spectra in the
\textit{ssp}, \textit{ppp} and \textit{sps} polarizations on the
air/water interface at four experimental configurations with
different incident angles for the visible and IR laser beams. They
are, Config.1: Visible=39$^{\circ}$, IR=55$^{\circ}$; Config.2:
Visible=45$^{\circ}$, IR=55$^{\circ}$; Config.3:
Visible=48$^{\circ}$, IR=57$^{\circ}$; Config.4:
Visible=63$^{\circ}$, IR=55$^{\circ}$.

\begin{table}[h!]
\caption{The fitting results of the SFG spectra at air/watrer
interface. The spectra are fitted with Lorentzian line shape
function as Eq.\ref{spectrum1}. The peak position of the
vibrational modes $\omega_{q}$, the peak width $\Gamma_{q}$ and
the oscillator strength factor $\chi_{eff,q,ijk}$ of the
vibrational modes are listed. The first column is the fitted value
for $\chi_{NR,eff,ijk}$. The relative error in fitting of
\textit{sps} is larger because of the small signal strength for
\textit{sps} spectra.}
\begin{center}
\begin{tabular}{lcccccccccccccc}
\hline  $\omega_{q}(cm^{-1})$ &   &              & 3281$\pm$5    & 3446$\pm$3    & 3536$\pm$6   & 3693$\pm$1& $$   \\
        $\Gamma_{q}(cm^{-1})$ &   &              & 89$\pm$9      & 103$\pm$7     & 77$\pm$11    & 17$\pm$1     \\
\hline
                     & ssp & 0.17  & -6.7$\pm$0.6  & -20.1$\pm$1.3 & -5.2$\pm$1.2 & 6.8$\pm$0.2  \\
Config.1             & ppp &-0.04  & 3.2$\pm$0.5   & 2.6$\pm$0.7   & 5.0$\pm$0.4  & 1.1$\pm$0.1  \\
                     & sps & -0.01  & -0.1$\pm$0.1  & -0.2$\pm$0.1   & -3.5$\pm$0.5 & 0.9$\pm$0.2 \\
\hline
                     & ssp & 0.19  & -8.3$\pm$0.6  & -24.0$\pm$3.5 & -5.0$\pm$3.5 & 8.5$\pm$0.1  \\
Config.2             & ppp &-0.02  & 1.1$\pm$0.5  & 0.0$\pm$0.8  & 6.9$\pm$0.3 & 2.4$\pm$0.6  \\
                     & sps & 0.02  & -0.1$\pm$0.1  & -0.3$\pm$0.1   & -4.5$\pm$0.5 & 1.6$\pm$0.1 \\
\hline
                     & ssp & 0.22  & -10.1$\pm$0.7 & -22.8$\pm$1.5 & -7.0$\pm$2.0  & 8.8$\pm$0.2  \\
Config.3             & ppp &-0.01  & 2.4$\pm$0.6  & 0.9$\pm$0.7  & 6.6$\pm$0.3 & 2.8$\pm$0.1  \\
                     & sps & 0.01  & -0.2$\pm$0.1  & -0.3$\pm$0.1   & -3.4$\pm$0.7 & 1.4$\pm$0.2 \\
\hline
                     &  ssp& 0.21  & -8.8$\pm$0.7  & -23.3$\pm$1.4 & -5.0$\pm$1.5 & 9.2$\pm$0.2  \\
Config.4             & ppp & 0.15  & -1.0$\pm$0.8  & -3.0$\pm$1.3  & -9.0$\pm$0.7 & 9.3$\pm$0.2  \\
                     & sps & 0.01  & -0.2$\pm$0.1  & -0.4$\pm$0.1   & -6.6$\pm$0.8 & 3.1$\pm$0.2 \\
\end{tabular}\label{fittingResults}
\end{center}
\end{table}

There are four apparent peaks can be identified in the SFG spectra
in Fig.\ref{allSpectra}. They are around $3700cm^{-1}$,
$3550cm^{-1}$, $3450cm^{-1}$ and $3250cm^{-1}$, respectively. The
$3700cm^{-1}$, $3450cm^{-1}$ and $3250cm^{-1}$ peaks has been
extensively discussed in the SFG
literature.\cite{Shen-prl1994,ShultzIRPC2000,
Richmond-jpca2000,richmond:jpcb1998,Shen-science} However, the
$3550cm^{-1}$ peak has been observed, but not yet clearly
identified or assigned.\cite{WeiXingPRL2001} The results of global
fit of these spectra with four Lorentzian peaks in
Eq.\ref{spectrum1} are listed in Table \ref{fittingResults}. From
the fitting results we can see that the peak bandwidths of the
$3550cm^{-1}$, $3450cm^{-1}$ and $3250cm^{-1}$ peaks are
$77\pm11cm^{-1}$, $103\pm 7cm^{-1}$ and $89\pm 9cm^{-1}$,
respectively. Such broad bandwidths indicate that they all belong
to different hydrogen bonded O-H stretching vibrational modes.
However, the bandwidth of the $3693cm^{-1}$ peak width is only
$17cm^{-1}$, consistent with the symmetric stretching (ss)
vibrational mode of the free O-H bond.\cite{DuQuanPRL1993} The
signs in Table \ref{fittingResults} contain the information of the
relative phase and interference effects of the different
vibrational modes. Here the phase of the $3693cm^{-1}$ peak is
held positive in each fit. Altering the relative phases of the
peaks on the same spectrum can not give a reasonable fit. Because
we used global fitting with all the spectra, these relative phases
can be determined accurately. They can be used to determine the
symmetry properties of each vibrational mode in Section IV.C.

\begin{figure}[h!]
\begin{center}
\includegraphics[height=5cm,width=6cm]{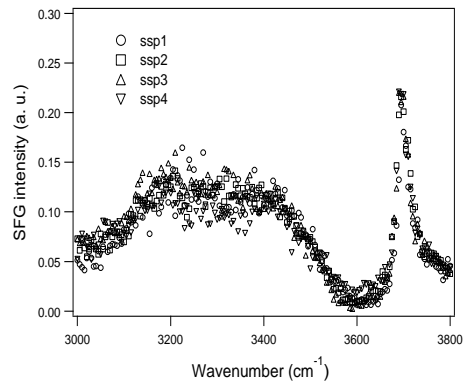}
\caption{Overlap of the normalized \textit{ssp} spectra of the
air/water interface in different experimental
configurations.}\label{TryOverlap}
\end{center}
\end{figure}

According to Eq.\ref{ppp}, the \textit{ssp} spectra in different
experimental configurations should have the same features from the
$\chi_{yyz}$ term. As shown in Fig.\ref{TryOverlap}, all
\textit{ssp} curves overlap quit well when normalized to the
$3693cm^{-1}$ peak. Calculation of the Fresnel factors with
different incident angles can quantitatively explain the relative
intensities in all four configurations.\cite{JohnsonJPCBpaper}
Because the SFG spectral intensity from the air/water interface in
the OH region is usually several times smaller than that of the
C-H region from other air/liquid interfaces, the air/water
interface SFG spectra are usually very hard to measure
experimentally. Therefore, the well overlapping of the
\textit{ssp} spectra in different experimental configurations is a
proof for the quality of our SFG-VS data. Furthermore, the spectra
we obtained agree very well with these in the
literatures.\cite{WeiXingPRL2001,AllenJPCB2004}

In principle, the \textit{sps} spectra in different experimental
configurations should also overlap with each other when
normalized. However, consistent with the calculations of the
corresponding Fresnel factors, the \textit{sps} signal level for
Config. 1, 2 and 3 are very close to the noise level, and features
in the \textit{sps} spectra can not be clearly identified except
for the spectra of Config.4. Therefore, such normalization and
comparison for \textit{sps} spectra is not as meaningful as the
\textit{ssp} spectra.

Different from the \textit{ssp} and \textit{sps} spectra, the
features in the \textit{ppp} spectra in Fig.\ref{allSpectra}
changed drastically with different experimental configurations.
This is because that the \textit{ppp} spectra is determined by
combination of four different $\chi_{ijk}$ tensors. Detailed
polarization analysis and experimental configuration analysis of
these changes in the \textit{ppp} spectra can provide symmetry
properties for each spectral features, as well as orientation and
structure information of the interfacial molecular groups, as
shall be shown later.\cite{HongfeiIRPCreview,Lurong2,Lurong3} We
shall show that analysis of the \textit{ppp} spectra in different
experimental configurations is very informative. However, this
advantage of \textit{ppp} spectra analysis has not been well
utilized in the previous literatures.

\subsection{Orientation and Motion of the Free OH Bond}\label{IVA}

Now with the knowledge of the SFG vibrational spectra of the
air/water interface, we can discuss the orientation and motion of
the free O-H bond at the air/water
interface.\cite{watershortpaper}

The sharp peak around $3700cm^{-1}$ was generally accepted as the
free OH bond protruding out of the liquid
water,\cite{DuQuanPRL1993,Richmond-jpca2000,WeiXingPRL2001,ShenLeePaper}
and it has been treated with $C_{\infty v}$ symmetry in
polarization analysis.\cite{DuQuanPRL1993,WeiXingPRL2001} Wei
\textit{et al.} studied the polarization dependence of the
intensity of this peak in the \textit{ssp}, \textit{ppp}, and
\textit{sps} polarizations measured with experimental
configuration of Visible$=45^{\circ}$,
IR=$57^{\circ}$.\cite{WeiXingPRL2001} Their SFG-VS data are
quantitatively very close to our data with Config.2 as expected.
Therefore, the \textit{ssp} intensity of the $3693cm^{-1}$ peak is
about 10 times of that of \textit{ppp}, and the \textit{sps}
intensity is essentially close zero. Wei \textit{et al.} realized
that using the step orientational distribution function in
Eq.\ref{StepAverage}, as well as other distribution functions,
such as Gaussian, centered at the surface normal, can not explain
such \textit{ssp}, \textit{ppp} and \textit{sps} intensity
relationships with the slow motion average in
Eq.\ref{SlowAverage}. On the other hand, the fast motion average
centered at the interface normal, as shown in Eq.\ref{FastAverage}
with $\theta_{M}=51^{\circ}$, can fairly well explain the observed
intensity relationships. Thus, it was concluded that the
orientation of the free OH bond of the interfacial water molecule
varies over a very broad angular range ($\theta_{M}=51^{\circ}$)
within the vibrational relaxation time as short as
$0.5ps$.\cite{WeiXingPRL2001}

\begin{eqnarray}
f(\theta)&=&cost\ \ \ \ for \ \ \ 0\leq \theta
\leq\theta_{M}\nonumber\\
f(\theta)&=&0\ \ \ \ \ \ \ \ for \ \ \ \theta \geq
\theta_{M}\label{StepAverage}
\end{eqnarray}

\begin{figure}[h!]
\begin{center}
\includegraphics[width=8cm]{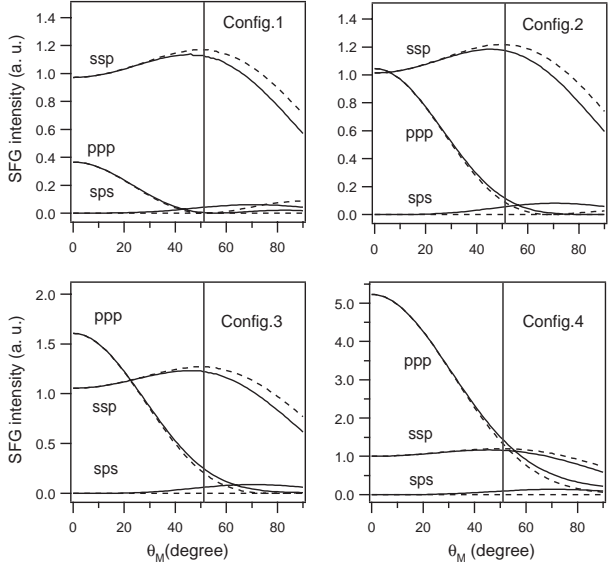}
\caption{SFG intensity of the free OH bond simulated with both
slow motion limit (solid curves) and fast motion limit (dotted
curves) following the procedure and parameters as Wei \textit{et
al.} \cite{WeiXingPRL2001}. $\theta_{M}$ is the range of
orientational motion of the free OH bond. All the curves presented
include the factor of $\sec^{2}\beta$, and all intensities are
normalized to the \textit{ssp} intensity in Config.4 with
$\theta_{M}=0^{\circ}$. The vertical lines indicate the
distribution width suggested by Wei \textit{et
al.}}\label{FourMotionSimulation}
\end{center}
\end{figure}

As shown in Fig.\ref{FourMotionSimulation}, Wei \textit{et al.}'s
treatment predicts clearly zero intensity for the \textit{sps}
spectra at $3693cm^{-1}$ with the assumption of fast orientational
motion centered at the surface normal. Using exactly the same
parameters, our calculation of Config.2 gives the same results as
that by Wei \textit{et al.} as it should have
been.\cite{WeiXingPRL2001} It is clear that the simulation results
in Fig.\ref{FourMotionSimulation} can fairly well explain the data
in Config. 1, 2 and 3, because all of them have relatively very
small \textit{sps} spectral intensity at $3693cm^{-1}$. However,
even though the fast orientational motion picture can explain the
relative intensity between the \textit{ssp} and \textit{ppp}
polarization in Config.4, it is clear that it can not explain the
apparently non-zero \textit{sps} intensity at $3693cm^{-1}$ with
Config.4. As long as the orientation distribution or orientational
motion is assumed to be centered to the interface
normal,\cite{ShenPrivate} orientational distribution functions
other than the step function in Eq.\ref{StepAverage} give the same
conclusion. Since the slow motion limit is already not an
option,\cite{WeiXingPRL2001} alternative description of the motion
and orientation at the air/water interface has to be invoked.

Because the air/water interface is rotationally isotropic around
the interface normal, now we assume that the molecular orientation
is centered around the tilt angle $\theta_{0}\neq 0$, instead of
the interface normal ($\theta_{0}=0$). If the Gaussian
distribution function is assumed, we have

\begin{eqnarray}
f(\theta)&=&\frac{1}{\sqrt{2\pi\sigma^{2}}}e^{-(\theta-\theta_{0})^{2}/2\sigma^{2}}\label{Gaussian}
\end{eqnarray}

\noindent in which $\sigma$ is the standard deviation of the
angular distribution. We shall show in the followings that by
using this distribution function, the $3693cm^{-1}$ peak in
different polarization and experimental configurations in Fig.
\ref{allSpectra} can be quantitatively analyzed.

\begin{table}[h!]
\caption{The general orientational parameter \textit{c} and the
strength factor \textit{d} for the vibrational stretching mode of
free OH bond in different experimental configurations. The
\textit{d} value bear the unit $\beta_{ccc}$ of single OH bond.}
\begin{center}
\begin{tabular}{lcccccccccccccc}
\hline  &  & d-ssp   &  c-ssp  &  d-sps &  c-sps &  d-ppp &  c-ppp  \\
\hline
Config.1  &  & 0.274  &   0.515  &   0.112  & 1  &   -0.154  &   1.53 \\
Config.2  &  & 0.256  &   0.515  &   0.118  & 1  &   -0.120  &   2.05 \\
Config.3  &  & 0.248  &   0.515  &   0.118  & 1  &   -0.104  &   2.43 \\
Config.4  &  & 0.176  &   0.515  &   0.107  & 1  &   -0.035  &   6.55 \\
\end{tabular}\label{CandDvalueForC3v}
\end{center}
\end{table}

Because the $3693cm^{-1}$ peak belongs to the \textit{ss} mode of
the free O-H bond at the air/water interface, it has been treated
with $C_{}\infty v$ symmetry. Now we calculate the general
orientational parameter \textit{c} and the strength factor
\textit{d} for the \textit{ssp}, \textit{sps} and \textit{ppp}
polarizations in all four experimental configurations with the
same parameters of the air/water interface as those used by Wei
\textit{et al.}\cite{WeiXingPRL2001} The details of the
calculation of $c$ and $d$ can be found
elsewhere.\cite{Lurong2,Lurong3,HongfeiIRPCreview} It is clear
from Table \ref{CandDvalueForC3v} that the $c$ values for the
\textit{ssp} and \textit{sps} polarizations are the same for all
four experimental configurations; whereas the $c$ values of the
\textit{ppp} polarization differ significantly for different
experimental configurations.

\begin{figure}[h!]
\begin{center}
\includegraphics[width=8cm]{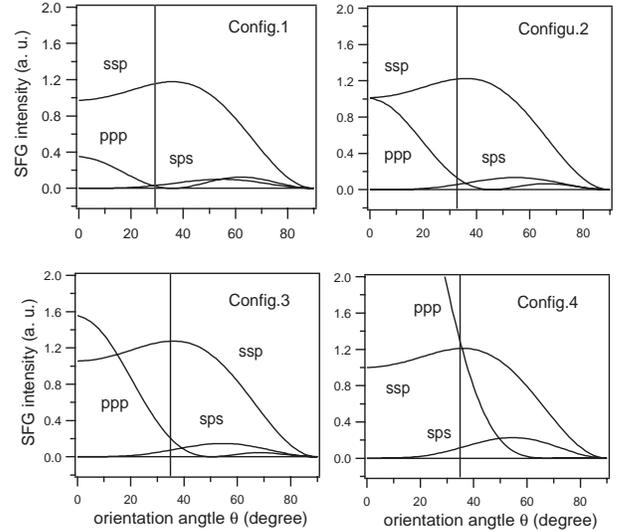}
\caption{The simulated SFG intensity of vibrational stretch mode
for free OH bond at different orientation angle $\theta$ assuming
$\sigma=0^{\circ}$. The factor $sec^{2}\beta$ in Eq.\ref{all} is
also included for comparison of SFG intensity in different
experimental configurations. All curves are normalized to the
\textit{ssp} intensity in Config.4 with $\theta_{0}=0^{\circ}$.
The vertical lines indicate the orientation which quantitatively
explains the observed SFG data.}\label{FourC3vSimulation}
\end{center}
\end{figure}

As we have demonstrated
previously,\cite{WHFRaoJCP2003,Lurong2,Lurong3,HongfeiIRPCreview}
the $[d\ast r(\theta)]^{2}$ vs. $\theta$ plot with $\sigma=0$ in
different polarizations can provide direct first look of the
physical picture for polarization analysis of SFG-VS data. Here we
plot $[d\ast r(\theta)\ast \sec\beta]^{2}$ vs. $\theta$ in
Fig.\ref{FourC3vSimulation} in order to compare data in different
experimental configurations. Thus, the relative intensity for the
$3693cm^{-1}$ peak in experimental Config.1, 2, 3 and 4 can be
used to calculate the orientation angle of the free O-H bond.
Using the known
procedures\cite{WHFRaoJCP2003,Lurong2,Lurong3,HongfeiIRPCreview}
and parameters,\cite{WeiXingPRL2001} they give the following four
values, i.e. $28.7\pm1.2^{\circ}$, $32.6\pm0.5^{\circ}$,
$34.6\pm0.7^{\circ}$ and $35.8\pm1.0^{\circ}$, respectively. These
values agree with each other quite well. However, the value from
Config.1, whose \textit{ppp} and \textit{sps} intensities are both
very weak, is not as reliable as the other three configurations.
Averaging over these values gives $\theta=33^{\circ}\pm
1^{\circ}$.

It is clear that $\sigma=0^{\circ}$ is not physically possible for
the liquid interface. However, the apparent success of the
quantitative explanation of the observed SFG spectra of the free
O-H bond in different experimental configurations using
$\sigma=0^{\circ}$ indicates that the actual $\sigma$ value can
not be very broad. Simulation of the $3693cm^{-1}$ peak in
different polarizations in each of the four experimental
configurations using the Gaussian distribution function in
Eq.\ref{Gaussian} concludes that $\sigma$ has to be smaller than
$15^{\circ}$ to satisfy the measured $3693cm^{-1}$ peak
intensities in all four experimental configurations.
$\sigma=15^{\circ}$ is the largest distribution width allowed by
the SFG experiment data for a Gaussian orientational distribution
function. With $\sigma=15^{\circ}$, we have
$\theta_{0}=30^{\circ}$. This indeed confirms that the orientation
of the free O-H bond is within a relatively narrow range (between
$30^{\circ}$ to $33^{\circ}$), with a relatively small
distribution width ($\sigma\leq 15^{\circ}$). Calculation with
both Eq.\ref{SlowAverage}, i.e. slow average limit, and
Eq.\ref{FastAverage}, i.e. fast average limit, gives
indistinguishable results with $\sigma$ as small as $\leq
15^{\circ}$ if $\theta_{0}$ is around $30^{\circ}$. This is
because that with a small distribution width, fast and slow motion
average should be the same according to Eq.\ref{FastAverage} and
Eq.\ref{SlowAverage}. Using a step distribution function around
$\theta_{0}\neq 0^{\circ}$ also give very close orientation angle
and distribution width.

Thus, our conclusion of the free O-H orientation and distribution
at the air/water interface is drastically different from the
conclusion given by Wei \textit{et al.}, which concluded that the
free O-H bond orientation varies in a broad range as big as
$102^{\circ}$ and as fast as $0.5$ picosecond, which is the
relaxation time for the O-H stretching
vibration.\cite{WeiXingPRL2001} It is clear that our conclusion is
based on the successful explanation of the observed polarized SFG
spectral intensities in different experimental configurations,
especially the relatively small but clearly non-vanishing SFG
spectral intensity at $3693cm^{-1}$ in the \textit{sps}
polarization. Our conclusion explicitly supports ultrafast
libratory motions with a relatively narrow angular range. As we
have known, the dynamics libratory motion of the hydrogen bonding
can be as fast as 0.1
picosecond.\cite{FeckoScience2003,Chandler1996PRL} Even with such
ultrafast dynamics, the air/water interface is nevertheless well
ordered. This is consistent with the generally well ordered
picture of the air/liquid and air/liquid mixture interfaces.
Recent quantitative analysis of data in SFG vibrational
spectroscopy have suggested that vapor/liquid interface are
generally well ordered, and sometimes even with anti-parallel
double-layered structures
\cite{ShenMirandaJPCBReview,ShenLinJCPAcetone2001,JohnsonJPCBpaper,ChenhuaJPCBacetone,
ChenhuaJPCBmethanol,ChenhuaCPLacetone}.

It has been generally accepted that liquid interface with strong
hydrogen bonding between molecules should be well ordered
\cite{ShenMirandaJPCBReview}. Our analysis here not only confirmed
this conclusion, but also provided solid and direct experimental
measurement of the orientation and motion at the air/water
interface.

\begin{figure*}[t]
\begin{center}
\includegraphics[width=7cm]{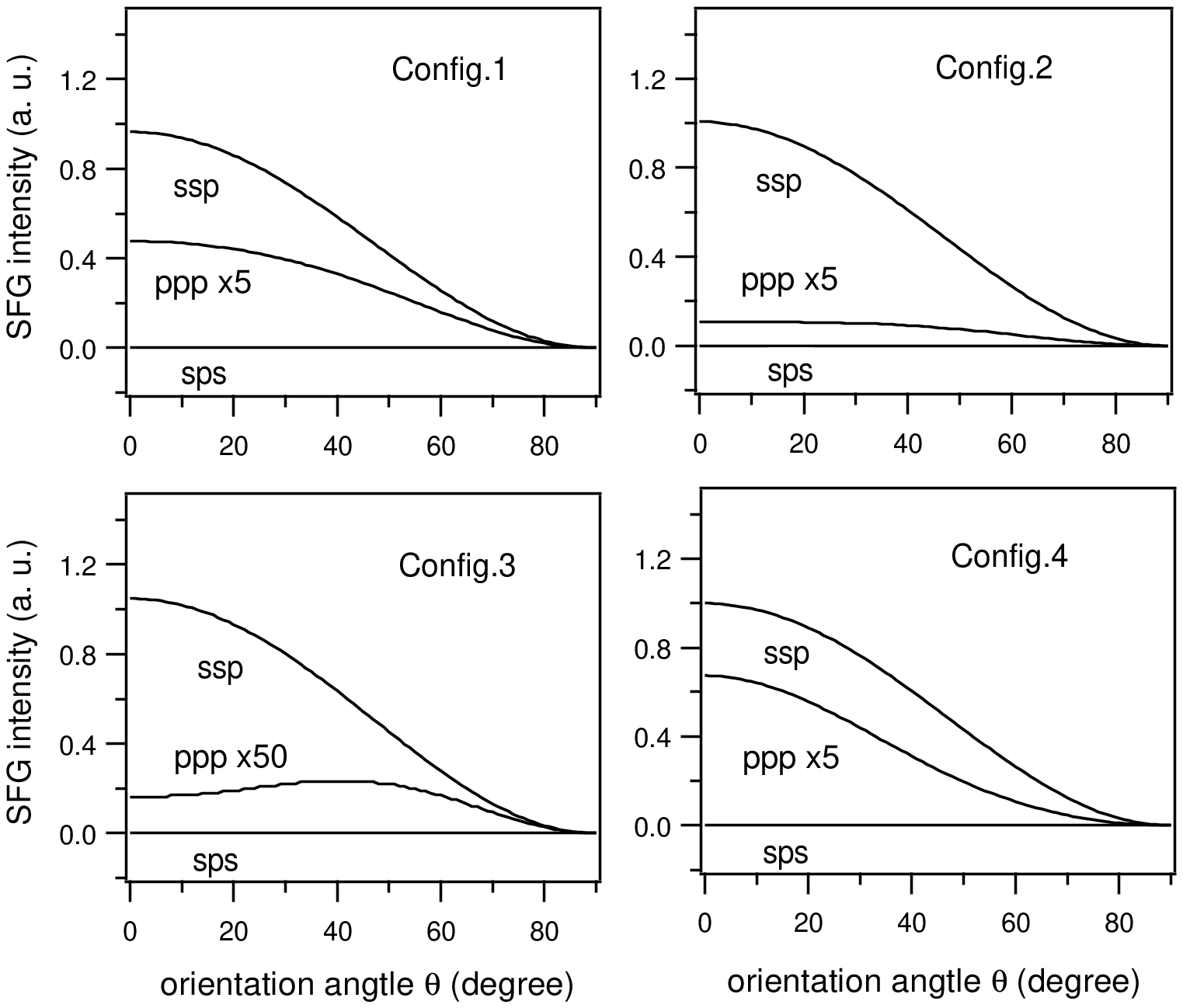}
\hspace{1.5cm}%
\includegraphics[width=7cm]{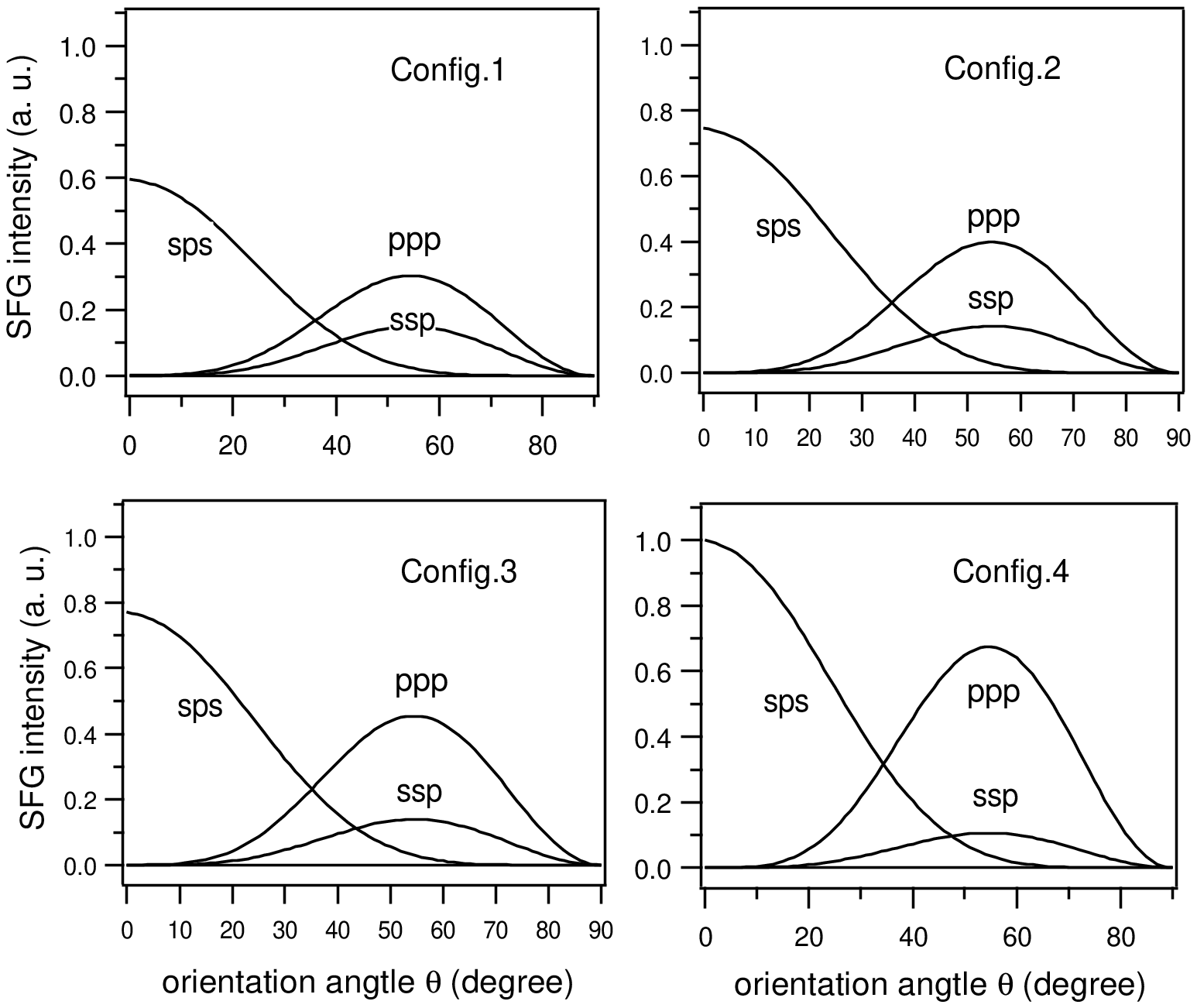}
\caption{The simulated SFG intensity for symmetric stretching
(\textit{ss}) mode (left) and asymmetric stretching (\textit{as})
mode (right) of water molecule with $C_{2v}$ symmetry. All curves
presented include the factor of $\sec^{2}\beta$. The intensity of
\textit{ss} mode is normalized to the \textit{ssp} intensity in
Config.4 with $\theta=0^{\circ}$. The intensity for \textit{as}
mode is normalized to the \textit{sps} intensity in Config.4 with
$\theta_{0}=0^{\circ}$. The units between plots of the \textit{ss}
and \textit{as}  modes differ by 9.11 times according to the
$\beta_{ccc}$ and $\beta_{aca}$ values in the
appendix.}\label{C2vsimulation}
\end{center}
\end{figure*}

\subsection{Polarization Analysis and Determination of Spectral Symmetry Property}

Here we try to apply polarization analysis for identifying the
symmetry property and for assignment of the SFG vibrational
spectra of the air/water interface.

The assignment of the SFG spectra of air/water interface in the
range of 3000 to 3800$cm^{-1}$ has been discussed
intensively.\cite{Shen-prl1994,RamanSpectroscopyBook,ShultzIRPC2000,
Richmond-jpca2000,richmond:science,RichmondCPL2004,
richmond:jpcb1998,Shen-science,richmondJPCB2003p546paper,RichmondARPC2001,Richmond:cr102:2693}
Richmond recently reviewed the current understanding of the
bonding and energetics, as well as the SFG spectra assignment, of
various aqueous interfaces, including the air/water
interface.\cite{RichmondARPC2001,Richmond:cr102:2693} The SFG
spectral assignments heavily relied on band fitting of IR and
Raman peak positions of bulk water or water cluster
spectra,\cite{Richmond:cr102:2693} as well as based on theoretical
calculations.\cite{TobiasJPCB2005,JPCA2000Buch,HynesCP2000,HynesJPCB2002}
The sharp peak at about 3700$cm^{-1}$ has been unanimously
assigned to the free O-H stretching vibration mode. The broad
peaks around 3250$cm^{-1}$ and 3450$cm^{-1}$ undoubtedly belong to
the hydrogen bonded O-H stretching modes, but their assignments
are not as unanimous as the 3700$cm^{-1}$ peak. The spectrum
around 3250$cm^{-1}$ was assigned to a continuum of O-H symmetric
stretches(ss), $\nu_{1}$ of water molecules in a symmetric
environment (ss-s), and was generally referred as "ice-like"
region because of its similarity in energy to O-H bonds in bulk
ice. The broad band around 3450$cm^{-1}$ was assigned to more
weakly correlated hydrogen bonded stretching modes, and was called
the "liquid-like" hydrogen-bonded region, where water molecules
reside in a more asymmetrically bonded (as) water
environment.\cite{RichmondARPC2001,Richmond:cr102:2693} The broad
peak around 3550$cm^{-1}$ appeared clearly in the \textit{ppp} SFG
spectra has been identified once and it has not been clearly
assigned so far.\cite{WeiXingPRL2001} Shultz \textit{et al.}
pointed out that these broad peak should also include the
asymmetric stretching mode of water molecules in a symmetry
environment and the bending overtone.\cite{ShultzIRPC2000}
Richmond \textit{et al.} also suggested that the intensity at
about 3450$cm^{-1}$ include the contribution of donor O-H
bond.\cite{richmond:science}

Recent progresses on SFG-VS have made it possible to determine the
symmetry properties of SFG-VS vibrational spectral features
through comparison of SFG spectra in different polarizations and
experimental
configurations.\cite{Lurong2,Lurong3,GanweiCPLNull,HongfeiIRPCreview}
The key idea of this development is from the commonsense of
molecular spectroscopy that vibrational modes of molecular groups
with different symmetry properties have different polarization
dependence on the interacting optical
fields.\cite{MichlBook,SymmetrySpectroscopyBook} Applying these
ideas to polarization analysis of SFG spectra has led to a set of
polarization selection rules for different stretching vibrational
modes of molecular groups with different molecular symmetry
properties, such as stretching vibrational modes of the $CH_{3}$
($C_{3v}$), $CH_{2}$ ($C_{2v}$) and $CH$ ($C_{\infty v}$)
groups.\cite{Lurong2,Lurong3,HongfeiIRPCreview} Many of these
selection rules are independent from molecular orientation and
orientational distribution. Therefore, they can be directly used
to identify symmetry property of SFG stretching vibrational band.
These progresses make it possible to analyze SFG vibrational
spectra \textit{in situ}, instead of rely only on the assignments
from Raman and IR studies of the bulk phases, which can be called
\textit{ex situ}. Because SFG spectra usually has more features
than those from IR and Raman measurement, some confusions and
errors in the previous spectral assignments have also been
clarified.\cite{Lurong2,Lurong3,HongfeiIRPCreview} Even though SFG
is naturally a polarized spectroscopy and the interfacial
molecular groups are usually ordered, this idea has not been
systematically explored until very
recently.\cite{WHFRaoJCP2003,Lurong2,Lurong3,HongfeiIRPCreview}

Water molecule possesses $C_{2v}$ symmetry. If the two O-H bond of
a water molecule are asymmetrically bonded, both O-H bond has to
be treated separately with $C_{\infty v}$ symmetry. This
classification of the water molecule symmetry is generally true no
matter it is hydrogen-bonded or not, in cluster, in bulk or at the
interface. Therefore, the symmetry property of the SFG vibrational
spectra features of the air/water interface can all be classified
accordingly. Thus, there are three kind of stretching vibrational
modes for us to deal with, namely, the symmetric (ss) and
asymmetric (as) stretching modes for $C_{2v}$ symmetry, and the
stretching mode for $C_{\infty v}$ symmetry. It is fairly easy to
distinguish these three stretching vibrational modes from the
polarization selection rules for SFG spectra of $CH_{2}$ and $CH$
groups.\cite{Lurong2,Lurong3,HongfeiIRPCreview} Because the bond
angle of the $CH_{2}$ group is slightly larger than that of water
molecule, the polarization dependence of the $C_{2v}$ water
molecule are slightly different. The key difference is that even
when the water molecule at the interface rotate freely around its
symmetry axis, the \textit{sps} spectral intensity of its
\textit{ss} mode does not vanish as that for $CH_{2}$. However,
this fact does not make the polarization selection rules different
for the interfacial $CH_{2}$ group and the $C_{2v}$ water
molecule.

Two of the major selection rules for the $C_{2v}$ group at a
dielectric interface are: \textit{(a) $\textit{ssp}$ intensity is
always many times of that of $\textit{ppp}$ for \textit{ss} mode.}
and \textit{(b) $\textit{ppp}$ intensity for \textit{as} mode is
always several times of that of $\textit{ssp}$. That is to say, if
there is any peak which is stronger in the $\textit{ssp}$ than
$\textit{ppp}$ spectra, it can not be from the $\textit{as}$
mode.}\cite{Lurong2,Lurong3} These two rules are independent from
molecular orientation and orientational distributions at a
rotationally isotropic interface.

It is clear from these selection rules, the sharp peak around
$3693cm^{-1}$ in Fig.\ref{allSpectra} does not belong to the
$C_{2v}$ symmetry, because its intensity in \textit{ppp}
polarization in Config. 1,2,3 is smaller than that in \textit{ssp}
polarization; while larger in Config.4. On the other hand, these
all fits well with the simulations in Fig.\ref{FourC3vSimulation}.
Therefore, $3693cm^{-1}$ peak is with $C_{\infty v}$ symmetry as
the free O-H stretching mode. Dissimilarly, both the broad peaks
around $3250cm^{-1}$ and $3450cm^{-1}$ in Fig.\ref{allSpectra} are
very strong only in the \textit{ssp} spectra in all experimental
configurations. They fit well with the \textit{ss} mode of the
$C_{2v}$ symmetry, and can not belong to the \textit{as} mode of
the $C_{2v}$ symmetry, or the $C_{\infty v}$ symmetry.

It is not so easy to determine the symmetry property of the broad
$3550cm^{-1}$ peak, because it appears to be buried in the high
frequency tail of the broad $3450cm^{-1}$ peak. It is not so
straight forward to read its relative intensity in \textit{ssp}
and \textit{ppp} polarizations from Fig.\ref{allSpectra}. However,
it appears significantly bigger in \textit{sps} polarization in
Config.4 than that in Config.1,2,3. Therefore, it appears to fit
with the simulations in Fig.\ref{FourC3vSimulation}. In order to
exclude the possibility that it may belong to a $C_{2v}$ mode (ss
or as), detailed simulation of the $C_{2v}$ modes in different
polarizations and experimental configurations is now called upon.

As describe in the appendix, the parameter \textit{c} and
\textit{d} of the $C_{2v}$ vibrational stretching modes are
calculated for different polarizations and experimental
configurations. Plots of $[d\ast r(\theta)\ast \sec\beta]^{2}$ vs.
tilted angle $\theta$ of the water molecule $c$ axis from the
interface normal using these $c$ and $d$ values are presented in
Fig.\ref{C2vsimulation}. These plots again confirm that the
$3693cm^{-1}$ can not belong to any $C_{2v}$ mode, especially with
the one order of magnitude increase of this peak in \textit{ppp}
polarization.

Clearly, the polarization dependence of the broad $3550cm^{-1}$
peak does not fit to the \textit{ss} mode of the $C_{2v}$
symmetry. Otherwise, according to Fig.\ref{C2vsimulation}, its
\textit{ppp} intensity in Config.3 has to be about one order of
magnitude weaker than that in the observed spectra. This peak can
not be the \textit{as} mode of the $C_{2v}$ symmetry either.
According to the \textit{c} and \textit{d} values for the
\textit{as} mode of the $C_{2v}$ symmetry, the phases in
\textit{ssp} and \textit{ppp} polarizations have to be with
opposite signs in all four experimental configurations. However,
fitting of the \textit{ssp} and \textit{ppp} spectra indicates
that in Config.4, the oscillator strengths had the same signs for
\textit{ssp} and \textit{ppp} polarizations, even though in
Congfig. 1,2,3, the oscillator strengths of this peak do possess
opposite phases these two polarizations. This indicates that the
$3550cm^{-1}$ peak is not $C_{2v}$ symmetry, and it appears to
have $C_{\infty v}$ symmetry.

Because in the bulk phase there is no observation of free O-H
bond, and because this broad peak $3550cm^{-1}$ appears to be
hydrogen-bonded, there is only one possibility that it is the
other O-H bond of the interfacial water molecule which has a free
O-H bond extruding away from the liquid bulk phase. According to
Fig.\ref{FourC3vSimulation}, the $C_{\infty v}$ O-H stretching
mode in \textit{sps} polarization is about twice as large in
Cogfig.4 as that in the other configurations. This is fully
consistent with the SFG spectra data in Fig.\ref{allSpectra} and
the fitting results in Table \ref{fittingResults}. Furthermore, in
Table \ref{fittingResults}, the phase of the broad $3550cm^{-1}$
peak is just opposite to that of the $3693cm^{-1}$ peak in both
\textit{ssp} and \textit{sps} polarizations, indicating these two
O-H pointing to opposite directions. The phase of the \textit{ppp}
polarization of the broad $3550cm^{-1}$ peak changes signs with
different experimental configuration. This is because the
orientational angle of the two O-H bonds are some times on the
same side of the minimum on the \textit{ppp} curves in
Fig.\ref{FourC3vSimulation}, and sometime on the different side of
the minimum, just as predicted with the experimental configuration
analysis. These detail features indicated the ability to
understand very subtle dependence of the SFG spectra on
experimental configurations and the parameters used for the
spectra calculations. Further study shall be reported elsewhere.

Further support for the assignment of the broad $3550cm^{-1}$ peak
came from the IR spectra measurement of the water dimer clusters,
where the stretching frequency for the donor O-H bond is just at
about $3550cm^{-1}$.\cite{RamanSpectroscopyBook,
PimentalWaterDimerPaper,NixonWaterDimerPaper,ShenLeePaper} This
assignment is a good support for our assignment of the peak at
3550$cm^{-1}$ in \textit{ppp} spectra to the single
hydrogen-bonded water molecule at the interface. Furthermore, the
two O-H stretching vibrations for the methanol dimer are at
3574$cm^{-1}$ and 3684$cm^{-1}$.\cite{MethanolJCP1991} The donor
O-H stretching mode is also in the same region of 3550$cm^{-1}$.

There is no observable spectra features in Fig.\ref{allSpectra}
for the \textit{as} mode of the $C_{2v}$ water molecules, neither
hydrogen bonded nor non-hydrogen bonded. According to
Fig.\ref{C2vsimulation}, for the \textit{as} modes corresponding
to the ss mode around $3250cm^{-1}$ and $3450cm^{-1}$, their
intensities have to be at least one order of magnitude weaker than
that of their corresponding \textit{ss} modes. It is
understandable that we do not observe them. Above discussion also
throw doubts on the existence of interfacial water molecules with
two free O-H bonds, as suggested somewhat less convincingly by
some recent studies.\cite{Richmond-jpca2000,richmond:science,
RichmondCPL2004,SaykallyJPCM2002,Mundy-science} According to the
polarization selection rules and the calculation for the
polarization dependence of the $C_{2v}$ water molecules, no
detectable spectral features satisfying the $C_{2v}$ symmetry in
the 3600$cm^{-1}$ and 3800$cm^{-1}$ has been observed in the SFG
spectra.

Here we clearly see that how polarization selection rules,
quantitative polarization and experimental configuration analysis
can help determine the symmetry property of the observed spectra
features. The importance of studying of spectral interference has
been demonstrated in recent
reports.\cite{Richmond-jpca2000,Shen2005PRLWaterQuartz,DaviesInterferenceJPCB2004}
Analysis in this work also demonstrated that, in order to discern
spectral details, it is useful and effective to analyze the
spectral interference of different spectral features through
global fitting of SFG spectra in different polarizations and
experimental configurations, and to compare fitting results with
the prediction from the calculated $c$ and $d$ values. This also
indicates the usefulness of the formulation of total SFG signal
with functions of $c$ and $d$ parameters in Eq.\ref{chi}.

\subsection{Molecular Structure at Air/Water Interface}

With the analysis of the orientation and motion, vibrational
spectral symmetry of the water molecules at the air/water
interface in above sections, we can have more understanding of the
molecular structure of the air/water interface.

In Section IV.B, we have determined that the free O-H oriented
around 30$^{\circ}$ away from the interface normal with a
orientational distribution narrower than $\sigma=15^{\circ}$, and
in Section IV.C, we have identified the spectral feature around
$3550cm^{-1}$ of the other hydrogen bonded O-H bond of this water
molecule. If the plane of this interfacial water molecule is close
to perpendicular to the interface, the orientation of the
hydrogen-bonded O-H should point into the liquid phase with a
orientation around 135$^{\circ}$ away from the interface normal.
This orientation is fully consistent with the calculation of the
polarization and experimental configuration dependence of the
broad $3550cm^{-1}$ peak with a $C_{\infty v}$ symmetry with the
observed SFG intensities, detail to be reported elsewhere. Such
orientation makes the dipole of this water molecule points around
97$^{\circ}$ from the interface normal. This picture is fully
consistent with conclusions in many previous experimental and
theoretical studies,
\cite{DuQuanPRL1993,Mundy-science,JaqamanJCP2004,LAAKSONENMolecularPhysics,
LAAKSONENJCP1997,TILDESLEYMolecularPhysics,HynesCP2000,GarrettJPC1996,RiceJCP1991,
BerkowitzCPL1991,WilsonJPC1987,TownsendJCP1985} but certainly
different from some.\cite{WeiXingPRL2001}

From Section IV.C, the broad spectral features between
3100$cm^{-1}$ to 3500$cm^{-1}$ are determined to be symmetric
stretching modes of the $C_{2v}$ symmetry. Because the peaks are
broad, and their energies is in the range of hydrogen-bonded O-H
stretching range, they can only come from the water molecules with
two donor O-H bonds, whose oxygen atom can accept either two, one
or zero hydrogen atom from other water molecules as hydrogen
donors. Certainly, the water molecule with the oxygen atom forming
two hydrogen-bonds is tetrahedral in shape and is "ice-like". This
is consistent with the previous assignment of the broad
$3250cm^{-1}$ peak.

The water molecule with no hydrogen bond for the oxygen atom is
obviously with $C_{2v}$ symmetry. However, the water molecule with
only one hydrogen bond for the oxygen atom may or may not preserve
the $C_{2v}$ symmetry. However, if this hydrogen bond perturbation
to the water structure is limited, this water molecule can still
be treated as with $C_{2v}$ symmetry. The last two kinds of water
molecules are certainly not "ice-like", but "liquid-like". The two
"liquid-like" species may have slightly different O-H vibrational
frequencies. However, only two apparently broad peaks in the
3100$cm^{-1}$ to 3500$cm^{-1}$ region have been identified in the
literatures.\cite{Shen-prl1994,RamanSpectroscopyBook,ShultzIRPC2000,
Richmond-jpca2000,richmond:science,RichmondCPL2004,
richmond:jpcb1998,Shen-science,richmondJPCB2003p546paper,RichmondARPC2001,Richmond:cr102:2693}
More studies on the possible hydrogen-bonded species are certainly
warranted in the future.

Here we confirm the conclusion by Brown \textit{et al.} that these
$C_{2v}$ water species all have their dipole vector point out of
the bulk liquid phase, i.e. with both hydrogen atoms point into
the bulk liquid phase.\cite{Richmond-jpca2000} It is clear in
Table \ref{fittingResults}, the signs of the \textit{ssp}
polarization oscillator strength factors of the $C_{2v}$ water
species are all in opposite phase to that of the free O-H peak at
$3693cm^{-1}$ in all experimental configurations. The signs and
values of the $c$ and $d$ parameters of the $C_{2v}$ and
$C_{\infty v}$ in Table \ref{CandDforC2vWater} and Table
\ref{CandDvalueForC3v}, respectively, indicate that the $c$ axis
of the $C_{2v}$ species has to be in opposite direction to the $c$
axis of the free O-H bond at the interface. Therefore, as defined
as in the appendix, the $C_{2v}$ species have to have their dipole
vector point out of the bulk liquid phase. The calculation of the
phase of the \textit{ppp} as well as \textit{sps} spectral
features are all consistent with this picture.

However, because the SFG spectral intensities of the \textit{ppp}
and \textit{sps} polarizations are generally in the noise level in
the 3100$cm^{-1}$ to 3500$cm^{-1}$ region (Fig.\ref{allSpectra}),
it is difficult to determine the range of the orientation angle
$\theta$ of these hydrogen-bonded species relative to the
interface normal. The orientational distribution of these $C_{2v}$
species can be quite broad, different from that for the
interfacial water molecules with the free O-H bond. From our
simulations, it appears to us that SFG measurement may not be very
effective to determine the orientational angle of the $C_{2v}$
species at the air/water interface, even though it can do very
well with the $C_{\infty v}$ O-H bonds as shown above. However,
our recent analysis of the SHG measurement of the neat air/water
interface showed that SHG measurement might be able to help
determine the orientational angle of the $C_{2v}$ species, but not
the $C_{\infty v}$ O-H bonds. Recent SHG results indicated that
the average orientation of the interfacial $C_{2v}$ water
molecules is about 40$^{\circ}$ to 50$^{\circ}$ from the surface
normal.\cite{wkzhangSHGwaterPaper}

The molecular structure, orientation and dynamics at nonpolar
material/water interfaces have been studied by \textit{ab initio}
calculation, MD simulation, or them
combined.\cite{Mundy-science,BenjaminPRL1994,HynesCP2000,
HynesJPCB2002,MooreJCP2003,ChandraCPL2003,ChandraCPL2004,RiceJCP1991,BenjaminCR1996,
JedlovskyWaterDCE,JedlovskyWaterCCl4} It appears that some
different conclusions were drawn on the molecular orientation and
structure of the air/water interface in different
studies.\cite{JedlovskyWaterCCl4,BenjaminPRL1994,MatsumotoJCP1987}
Nevertheless, many of these studies concluded that the dipole
vector of the interfacial water molecules prefers lying parallel
to the interface and have one of the O-H bond protrude out of the
liquid phase. The majority of the conclusions from theoretical
calculations agree satisfactorily with the experimental analysis
of ours and previous studies, but all the simulation results were
with significantly broader orientational
distributions.\cite{Mundy-science,JaqamanJCP2004,LAAKSONENMolecularPhysics,
LAAKSONENJCP1997,TILDESLEYMolecularPhysics,HynesCP2000,GarrettJPC1996,RiceJCP1991,
BerkowitzCPL1991,WilsonJPC1987,TownsendJCP1985} There were reports
concluded that some interfacial water molecules have their two O-H
bonds projecting into the vapor phase and with oxygen atoms in the
liquid phase.\cite{Mundy-science,GrayCondencedMatter1994,
RobinsonJPC1991,CroctonPhysica1981} However, we have not found
explicit spectroscopic evidence for such species at the air/water
interface. These all indicate that detailed comparison of the
theoretical calculations and the experimental analysis is
certainly an important subject in the future studies.

\section{Conclusion}

Detailed understanding of the air/water interface is important,
and can be used for the general understanding of the liquid water
structure. In this work, we presented detailed analysis of the SFG
vibrational spectra of the air/water interface taken in different
polarizations and experimental configurations. Polarization and
experimental configuration analysis have provided detailed
information on the orientation, structure and dynamics of the
water molecules at the air/water interface. The success of these
analysis indicated the effectiveness and ability of SFG-VS as a
uniquely interface specific spectroscopic probe of liquid
interfaces and other molecular interfaces. It also indicates that
for the neat air/water interface, as has been studied in the
literature for some other simple air/liquid interfaces, the
contribution from the interface region dominates the SFG
spectra.\cite{ShenApplyPhys,WeiXinJPCBBulkVsSurface,WeiPRBBulkvsSurface,Shen-methanolPRL1991}

Here are major conclusions we have reached for the air/water
interface. Firstly, we concluded that the motion of the
interfacial water molecules can only be in a limited angular
range, instead rapidly varying over a broad angular range in the
vibrational relaxation time suggested previously. Secondly,
because different vibrational modes of different molecular species
at the interface has different symmetry properties, polarization
and symmetry analysis of the SFG-VS spectral features can help
assignment of the SFG-VS spectra peaks to different interfacial
species. These analysis concluded that the narrow $3693cm^{-1}$
and broad $3550cm^{-1}$ peaks belong to $C_{\infty v}$ symmetry,
while the broad $3250cm^{-1}$ and $3450cm^{-1}$ peaks belong to
the symmetric stretching modes with $C_{2v}$ symmetry. Thus, the
$3693cm^{-1}$ peak is assigned to the free OH, the $3550cm^{-1}$
peak is assigned to the single hydrogen bonded OH stretching mode,
and the $3250cm^{-1}$ and $3450cm^{-1}$ peaks are assigned to
interfacial water molecules as two hydrogen donors for hydrogen
bonding (with $C_{2v}$ symmetry), respectively. Thirdly, analysis
of the SFG-VS spectra concluded that the singly hydrogen bonded
water molecules at the air/water interface have their dipole
vector direct almost parallel to the interface, and is with a very
narrow orientational distribution. The doubly hydrogen bond donor
water molecules have their dipole vector point away from the
liquid phase. Finally, we did not find any observable evidence for
interfacial water molecules with doubly free O-H bonds at the
air/water interface.

Many of the conclusions in this work agree well with previous
reports, with much more detailed understandings. The conclusion of
the narrow range motion of the free O-H bond is different from the
literature. The explicit assignment of the broad $3550cm^{-1}$
peak and determination of the symmetry property of the
hydrogen-bonded O-H stretching modes in the 3100$cm^{-1}$ to
3500$cm^{-1}$ region are based on firm evidences. These
conclusions as a whole provided a detailed and general picture of
the spectroscopy, structure and dynamics of the air/water
interface, which can be used for understanding chemical and
biological problems related to the ubiquitous water molecule in
general. The concepts and approaches used in the analysis in this
report can be applied to studying on more complex molecular
interfaces.

Recently, extensive efforts with SFG-VS, as well as SHG,
experimental studies and theoretical simulations have been devoted
to the renewed interests on ion adsorption and the Jones-Ray
effect at the air/aqueous solution
interfaces.\cite{TobiasJPCB2005,ShultzIRPC2000,RichmondJPCB2004,ShultzJPCB2002,
AllenJPCB2004,SaykallyCPL2004,SaykallyCPL2004-2,SaykallyJPCB2005,MuchaJPCB2005}
We suggest that detailed polarization and experimental
configuration analysis of the SFG vibrational spectra be applied
to these interfaces.

\vspace{0.8cm}

\noindent \textbf{Acknowledgment.} This work was supported by the
Chinese Academy of Sciences (CAS, No.CMS-CX200305), the Natural
Science Foundation of China (NSFC, No.20425309) and the Chinese
Ministry of Science and Technology (MOST, No.G1999075305). We
thank Bao-hua Wu for help derive the bond polarizability
derivative model expressions. H.F.W. acknowledges Y. R. Shen for
helpful discussions.

\section*{Appendix: Calculation of \textit{d} and \textit{c} Parameters for $C_{2v}$ Molecule}

Here we present the expressions to calculate the parameter
\textit{c} and \textit{d} for water molecules with C$_{2v}$
symmetry using the bond polarizability derivative model first used
by Hirose \textit{et al}.\cite{hirose:jcp1992,hirose:jpc1993} The
detailed re-derivation of the complete expressions and the
effectiveness of the model can be found in a recent
review.\cite{HongfeiIRPCreview}

The relationship between the Raman depolarization ratio $\rho$ and
the bond polarizability \textit{r} for a molecule group with
C$_{2v}$ symmetry was:\cite{HongfeiIRPCreview}

\begin{eqnarray}
\rho&=\frac{3}{4+20\frac{(1+2r)^2}{(1-r)^2(1+3\cos^2\tau)}}\label{rCH2definition}
\end{eqnarray}

\noindent in which $\tau$ is the H-O-H bond angle between the two
OH bonds of a water molecule. With the Raman depolarization ratio
measured as about 0.03,\cite{Murphy-r} the bond polarizability
\textit{r} for OH bond in water molecule can be deduced to be
0.32, as used by Du \textit{et al.}\cite{DuQuanPRL1993}

The 7 hyperpolarizability tensor elements of water molecule with
$C_{2v}$ symmetry are as the followings.

\begin{eqnarray}
\nonumber\beta_{aac}&=&\frac{G_{a}\beta_{OH}^{0}}{\omega_{a1}}\left[(1+r)-(1-r)\cos\tau\right]\cos(\frac{\tau}{2})\\
\nonumber\beta_{bbc}&=&\frac{2G_{a}\beta_{OH}^{0}}{\omega_{a1}}r\cos(\frac{\tau}{2})\\
\nonumber\beta_{ccc}&=&\frac{G_{a}\beta_{OH}^{0}}{\omega_{a1}}\left[(1+r)+(1-r)\cos\tau\right]\cos(\frac{\tau}{2})\\
\nonumber\beta_{aca}&=&\beta_{caa}=\frac{G_{b}\beta_{OH}^{0}}{\omega_{b1}}\left[(1-r)\sin\tau\right]\sin(\frac{\tau}{2})\\
\beta_{bcb}&=&\beta_{cbb}=0\label{c2vbond}
\end{eqnarray}

\noindent Where G$_{a}=(1+\cos\tau)/M_{O}+1/M_{H}$ and
G$_{b}=(1-\cos\tau)/M_{O}+1/M_{H}$ are the inverse effective mass
for the symmetric ($a_{1}$) and asymmetric ($b_{1}$) normal modes,
with $M_{O}$ and $M_{H}$ as the atomic mass of O and H atoms,
respectively. $\omega_{a1}$ and $\omega_{b1}$ are the vibrational
frequencies of the respective modes.
$\beta_{OH}^{0}=\frac{1}{2\varepsilon_{0}}\alpha'_{\zeta\zeta}\mu'_{\zeta}$,
as defined by Wei \textit{et al.}\cite{weixing:pre2000} The water
molecule are fixed in the molecular coordination $\lambda'(a,b,c)$
with the O atom at the coordination center, the molecule plane in
\textit{ac} plane, and the bisector from the oxygen to the two
hydrogen atoms side is the \textit{c} axis.

For the achiral rotationally isotropic ($C_{\infty v}$) liquid
interface, the symmetric stretching (\textit{ss}, $a_{1}$)
vibrational modes have,\cite{HongfeiIRPCreview}

\begin{eqnarray}
\chi_{xxz}^{(2),ss}&=&\chi_{yyz}^{(2),ss}\nonumber\\
&=&\frac{1}{2}N_{s}[\langle\cos^{2}\psi\rangle\beta_{aac}+
\langle\sin^{2}\psi\rangle\beta_{bbc}+\beta_{ccc}]\langle\cos\theta\rangle\nonumber\\
&+&\frac{1}{2}N_{s}[\langle\sin^{2}\psi\rangle\beta_{aac}+\langle\cos^{2}\psi\rangle\beta_{bbc}-
\beta_{ccc}]\langle\cos^{3}\theta\rangle\nonumber\\
\chi_{xzx}^{(2),ss}&=&\chi_{zxx}^{(2),ss}=\chi_{yzy}^{(2),ss}=\chi_{zyy}^{(2),ss}\nonumber\\
&=&-\frac{1}{2}N_{s}[\langle\cos\theta\rangle\nonumber-\langle\cos^{3}\theta\rangle]\nonumber\\
&&[\langle\sin^{2}\psi\rangle\beta_{aac}+\langle\cos^{2}\psi\rangle\beta_{bbc}
-\beta_{ccc}]\nonumber\\
\chi_{zzz}^{(2),ss}&=&N_{s}[\langle\sin^{2}\psi\rangle\beta_{aac}+\langle\cos^{2}\psi\rangle\beta_{bbc}]
\langle\cos\theta\rangle\nonumber\\
&-&N_{s}[\langle\sin^{2}\psi\rangle\beta_{aac}+\langle\cos^{2}\psi\rangle\beta_{bbc}-
\beta_{ccc}]\langle\cos^{3}\theta\rangle\nonumber\\
\label{ssofC2Vpsi}
\end{eqnarray}

\noindent And the asymmetric stretching (\textit{as}, $b_{1}$)
vibrational modes have,

\begin{eqnarray}
\chi_{xxz}^{(2),as}&=&\chi_{yyz}^{(2),as}=-N_{s}\beta_{aca}\langle\sin^{2}\psi\rangle
[\langle\cos\theta\rangle-\langle\cos^{3}\theta\rangle]\nonumber\\
\chi_{xzx}^{(2),as}&=&\chi_{zxx}^{(2),as}=\chi_{yzy}^{(2),as}=\chi_{zyy}^{(2),as}\nonumber\\
&=&\frac{1}{2}N_{s}\beta_{aca}[\langle\cos^{2}\psi\rangle-\langle\sin^{2}\psi\rangle]
\langle\cos\theta\rangle\nonumber\\
&+&N_{s}\beta_{aca}\langle\sin^{2}\psi\rangle\langle\cos^{3}\theta\rangle\nonumber\\
\chi_{zzz}^{(2),as}&=&2N_{s}\beta_{aca}\langle\sin^{2}\psi\rangle
[\langle\cos\theta\rangle-\langle\cos^{3}\theta\rangle]\label{asofC2Vpsi}
\end{eqnarray}

The $b_{2}$ asymmetric mode are SFG inactive since the
hyperpolarizability tensors $\beta_{bcb}$ and $\beta_{cbb}$ are
zero.

The Euler angel $\psi$ can be integrated if the H-X-H plane of the
XH$_{2}$ group can rotate freely around its symmetry axis $c$. For
water molecules with both OH bond hydrogen bonded to neighboring
molecules in liquid phase, the Euler angel $\psi$ should not be a
fixed value. Assuming a random $\psi$ distribution we have the
following non-vanishing tensor elements for the
symmetric-stretching mode.,\cite{Lurong2,HongfeiIRPCreview}

\begin{eqnarray}
\chi_{xxz}^{(2),ss}&=&\chi_{yyz}^{(2),ss}=\frac{1}{4}N_{s}(\beta_{aac}
+\beta_{bbc}+2\beta_{ccc})\langle{cos\theta}\rangle\nonumber\\
&+&\frac{1}{4}N_{s}(\beta_{aac}+\beta_{bbc}-2\beta_{ccc})
\langle{cos^3\theta}\rangle\nonumber\\
\chi_{xzx}^{(2),ss}&=&\chi_{zxx}^{(2),ss}=\chi_{yzy}^{(2),ss}=\chi_{zyy}^{(2),ss}\nonumber\\
&=&-\frac{1}{4}N_{s}(\beta_{aac}+\beta_{bbc}-2\beta_{ccc})
(\langle{cos\theta}\rangle-\langle{cos^3\theta}\rangle)\nonumber
\\\chi_{zzz}^{(2),ss}&=&\frac{1}{2}N_{s}(\beta_{aac}+\beta_{bbc})
\langle{cos\theta}\rangle\nonumber\\
&-&\frac{1}{2}N_{s}(\beta_{aac}+\beta_{bbc}
-2\beta_{ccc})\langle{cos^3\theta}\rangle\label{ssofC2V}
\end{eqnarray}

\noindent And the non-vanishing tensor elements for water
asymmetric-stretching modes are,

\begin{eqnarray}
\chi_{xxz}^{(2),as}&=&\chi_{yyz}^{(2),as}=-\frac{1}{2}N_{s}\beta_{aca}
(\langle{cos\theta}\rangle-\langle{cos^3\theta}\rangle)\nonumber\\
\chi_{xzx}^{(2),as}&=&\chi_{zxx}^{(2),as}=\chi_{yzy}^{(2),as}=\chi_{zyy}^{(2),as}
=\frac{1}{2}N_{s}\beta_{aca}\langle{cos^3\theta\rangle}\nonumber\\
\chi_{zzz}^{(2),as}&=&
N_{s}\beta_{aca}(\langle{cos\theta}\rangle-\langle{cos^3\theta\rangle})\label{asofC2V}
\end{eqnarray}

\begin{table}[t]
\caption{The general orientational parameter \textit{c} and the
strength factor \textit{d} for \textit{ss} mode and \textit{as}
mode of water molecule with $C_{2v}$ symmetry in different
polarization combinations. The \textit{d} values bear the unit
$\beta_{ccc}$.}
\begin{center}
\begin{tabular}{lcccccccccccccc}
\hline
\textit{ss} mode&  & d-ssp   &  c-ssp  &  d-sps &  c-sps &  d-ppp &  c-ppp  \\
\hline
Config.1        &  & 0.400  &   0.038  &   0.012  & 1  &   -0.146  &   0.174 \\
Config.2        &  & 0.374  &   0.038  &   0.013  & 1  &   -0.079  &   0.338 \\
Config.3        &  & 0.362  &   0.038  &   0.013  & 1  &   -0.046  &   0.589 \\
Config.4        &  & 0.257  &   0.038  &   0.012  & 1  &    0.066  &   -0.378 \\
\hline
\textit{as} mode&  & d-ssp   &  c-ssp  &  d$\ast$c-sps &  c-sps &  d-ppp &  c-ppp \\
\hline
Config.1        &  & -0.154  &   1  &   -0.122  & $\infty$  &   0.262  &   0.98 \\
Config.2        &  & -0.144  &   1  &   -0.129  & $\infty$  &   0.272  &   0.99 \\
Config.3        &  & -0.139  &   1  &   -0.128  & $\infty$  &   0.277  &   0.99 \\
Config.4        &  & -0.099  &   1  &   -0.117  & $\infty$  &   0.250  &   1.01 \\
\end{tabular}\label{CandDforC2vWater}
\end{center}
\end{table}

For CH$_{2}$ group, there is a general relationship
$\beta_{aac}+\beta_{bbc}-2\beta_{ccc}\cong 0$, because $\tau
=109.5^{\circ}$.\cite{Lurong2,HongfeiIRPCreview} This relationship
makes $\chi_{xzx}^{(2),ss}$=$\chi_{zxx}^{(2),ss}$=
$\chi_{yzy}^{(2),ss}$=$\chi_{zyy}^{(2),ss}\cong 0$, which means
that the \textit{ss} vibrational mode should vanish in the
\textit{sps} and \textit{pss} polarizations according to Eq.
\ref{ssofC2V}. For water molecule, $\tau=104.5^{\circ}$. Then
hyperpolarizability tensors of the water molecule are as the
followings: $\beta_{aac}=1.296$; $\beta_{bbc}=0.557$;
$\beta_{ccc}=1$; $\beta_{aca}=\beta_{caa}=0.741$;
$\beta_{bcb}=\beta_{cbb}=0$. Here all value are normalized to
$\beta_{ccc}=1$. Then,
$\beta_{aac}+\beta_{bbc}-2\beta_{ccc}=-0.147$. This value is not
$0$, but is very small. So the \textit{ss} vibrational mode
spectra in the \textit{sps} and \textit{pss} polarizations should
vanish as the $CH_{2}$ group mentioned above. However, they have
to be very small comparing with in \textit{ssp} spectra. This is
fully consistent with the small intensities in the \textit{sps}
SFG spectra for the $C_{2v}$ water modes in Fig.\ref{allSpectra}.

With above deduction, and following the procedure in previous
report,\cite{Lurong3} the general orientational parameter
\textit{c} and strength factor \textit{d} for the symmetric
stretching (\textit{ss}) mode and asymmetric stretching
(\textit{as}) mode of water molecule in different polarizations
and experimental configurations can be calculated (see Table
\ref{CandDforC2vWater}). The parameters used in the calculation
are $n_{1}(\omega)$=$n_{1}(\omega_{1})$=$n_{1}(\omega_{2})$=1;
$n_{2}(\omega)$=$n_{2}(\omega_{1})$=1.34;
$n_{2}(\omega_{2})$=1.18; $n'(\omega)$=$n'(\omega_{1})$=1.15;
$n'(\omega_{2})$=1.09, respectively. These parameters are the same
as the dielectric constants used for calculation of the air/water
interface by Wei \textit{et al.}\cite{WeiXingPRL2001} As we have
discussed in our reports,\cite{Lurong2,Lurong3} polarization
analysis with the co-propagating experimental geometry is
insensitive to the value of the dielectric constants of the IR
frequency.\cite{GanweiCPLNull,HongfeiIRPCreview} Therefore, we
used the same refractive constants for the IR frequencies across
the whole $3100cm^{-1}$ to $3800cm^{-1}$ region, and this does not
appear to affect our analysis. These \textit{c} and \textit{d}
values are used to calculate the polarization and orientation
dependence of the SFG intensity, as well as the interference
(phase) of different spectral features in different experimental
configurations. These calculations can satisfactorily explain the
detailed changes of the observed spectral features, as discussed
in the main text.

It is to be noticed that in the above discussion we only used
single water molecule parameters. When there is association and
clustering of water molecules, as long as the $C_{2v}$ symmetry
preserves, and the H-O-H bond angle does not change significantly,
above expressions dictated by symmetry properties should still be
valid.

\end{document}